\documentclass{article}
\usepackage[latin9]{inputenc}
\usepackage{geometry}
\usepackage{amsmath}
\usepackage{amssymb}
\usepackage[unicode=true,pdfusetitle,
 bookmarks=true,bookmarksnumbered=false,bookmarksopen=false,
 breaklinks=false,pdfborder={0 0 0},backref=section,colorlinks=false]
 {hyperref}

\makeatletter

\usepackage{float}
\usepackage{graphicx}
\usepackage{grffile}
\usepackage{breakurl}
\usepackage{caption}

\usepackage{fullpage}\usepackage[justification=centering]{caption}\usepackage[all]{hypcap}\usepackage{microtype}

\makeatother

\begin{document}

\title{Manifestations of Isospin in Nearest Neighbor Spacing Distributions for the f-p Model Space}

\author{Michael Quinonez, Arun Kingan and Larry Zamick\\
\\
 \textit{Department of Physics and Astronomy}, Rutgers University,
Piscataway, New Jersey 08854 }

\date{\today}
\maketitle
\begin{abstract}
The strong interactions are charge independent. If we limit ourselves
to the strong interactions, we have the isospin $T$ as a good quantum number. Here we consider the
lack of level repulsion of states of different isospin and how this effect manifests in nearest neighbor spacing (NNS) histograms, which provide a visual and statistical context in which to study distributions of energy level spacings. In particular, we study nucleons in the f-p model space for the nucleus $^{44}$Ti. We also study the effect  of the Coulomb interaction on the level spacing distribution.
\end{abstract}

\section{Introduction.}

If we limit ourselves to charge independent interactions, e.g. the
strong interactions, we can classify nuclear ground and excited
states by the isospin quantum number $T$. The neutron has $T=1/2$ and projection $T_{z}=1/2$
whilst the proton has $T=1/2$, $T_{z}=-1/2$. A given nucleus has $T_{z}
= (N-Z)/2$. We have the rule $T \geq T_{z}$, and that an isospin
$T$ corresponds to a multiplet with $(2T+1)$ members. For example, in the single j shell model of $^{44}$Ti, the 2 valence protons and 2 valence neutrons are in the
f$_{7/2}$ shell. In this model one can form states of isopin $T=$ 0, 1, 
and 2. The $T=0$ states are isosinglet i.e. they only occur in $^{44}$Ti. For $T=1$ in $^{44}$Ti there are analogs in $^{44}$Sc and $^{44}$V.
For T=2 the multiplet members, all with A=44, are Ca, Sc, Ti, V, and Cr.
For non-zero $T$ if one knows the wave function of a state $(T,T_{z})$
then acting with the lowering operator $T_{-}$ one obtains the wave
function of the state $(T,T_{z-1})$ in a neighboring nucleus. The
$T_{+}$ operator will take us to the state $(T, T_{z+1})$.

The matrix element of a charge independent interaction between 2 states
of different isospin will be zero - hence no level repulsion arises. This has
dramatic effects on NNS distributions as will be discussed in the next section.

\section{Nearest neighbor spacings in $^{44}$Ti.}

A nearest neighbor spacing histogram depicts the behavior of spacings between adjacent elements in a list of numbers. In the context of nuclear energy level spacings, one can produce an NNS histogram by generating a list of energy levels (i.e. from experiment or shell model calculations) and taking the difference between each energy level and the one which immediately precedes it. These spacings are converted into units of their mean spacing, sorted into groups which fall between certain spacing intervals, and plotted such that each interval on the abscissa is assigned a bar whose height along the ordinate is proportional to the number of spacings in that interval. The final NNS histogram is a probability density plot, in which the area of a particular bar gives the probability of choosing a spacing from that particular interval at random.

 If one gets energy levels from a random matrix then an NNS histogram is described by a Poisson (or exponential) distribution. If one uses instead a matrix Hamiltonian for a many nucleon system derived from a reasonable nucleon-nucleon
interaction then one also gets a Poisson distribution. This distribution has the property that its mean is equal to its variance, which gives a quantitative way to check for Poisson behavior. However, Wigner
{[}1,2{]} realized that level repulsion should somehow come into play.
In a Poisson distribution the peak is at zero level spacing, which,
at first glance would suggest that level repulsion is not important.
However Wigner realized that the small spacings could be due to certain
symmetries{[}1{]}. For example, states of different total angular
momentum do not mix. Wigner obtained the following distribution when
states of only one symmetry were included:
\begin{equation}
P(s) = \frac{\pi}{2} s \cdot \exp \left( -\frac{\pi}{4}s^2 \right)
\end{equation}
where $s$ is the level spacing divided
by the mean level spacing and $P(s)$ is the probability density function of $s$. It is normalized so that the integral over
all positive $s$ is one. It vanishes at $s=0$ in contrast to the Poisson
distribution, which has the form
\begin{equation}
 P(s) = \exp(-s)
\end{equation} 
  Wigner's work has stimulated many other works
on level densities and spacings. Here we list a few {[}2-10{]}.

In this work, in the spirit of Wigner, we will also consider the
effects of symmetries on NNS distributions.
We do so in part by mixing and then unmixing states of different angular
momentum, simply as a demonstration of level repulsion, but our main focus will be on isospin symmetry. With a charge
independent interaction the matrix elements of basis states of different
isospin vanish. We will first consider $T=0$, 1, and 2 states of angular
momentum $J=4^{+}$. We show in Figure 1 the NNS distributions
in $^{44}$Ti for these $J=4^{+}$ states resulting from a shell model matrix diagonalization with the
Nushellx program of B.A. Brown and W..M. Rae {[}11{]}. We use the GXFP1 interaction in the f-p model space.
Our variable is the nearest neighbor level spacing divided by the
mean level spacing. First we have $T=0$ only, $T=1$
only and $T=2$ only. Then we have $T=0$ mixed with either $T=1$ or $T=2$, and finally we mix all isospins. This 
is repeated for $J=5^{+}$ in Figure 2. We then show a large mixture of all isospins and both $J=4^+$ and $J=5^+$ as a demonstration of further level repulsion in Figure 3. Furthermore, in the interest of comparing energy spacing distributions in isospin formalism to those in proton-neutron formalism, we include histograms produced by including the Coulomb interaction. In each figure we overlay the Poisson and Wigner distributions.

We omit the ground state region (lowest 5-10$\%$ of spacings in each $J$ and $T$ configuration) from these studies due to the comparably large spacings in this region and their effects on the variances of each set of spacings. Even a few of such large spacings (between 5-30 times the mean level spacing) can more than quintuple the variance of a sample by virtue of the dominance of their corresponding terms in the variance sum $\sigma ^2=\sum_{i}^N (X_i - \mu)^2/N$, where $X_i$ is a particular element in the sample, $\mu$ is the mean, and $N$ is the number of elements in the sample. Since these spacings are far outnumbered by those closer to the mean spacing in our samples, the variances we obtain by removing the ground state region provide a more accurate description of the general behavior of our distributions. 

\begin{figure}[H]
\caption{NNS histograms for $J=4^+$ states with various $T$}
\includegraphics[scale=0.43]{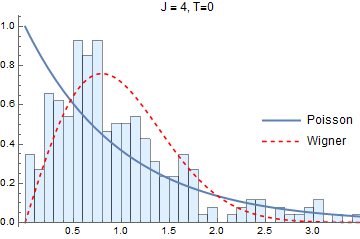}
\includegraphics[scale=0.43]{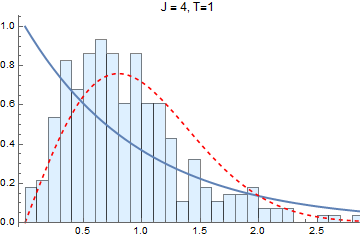}
\includegraphics[scale=0.43]{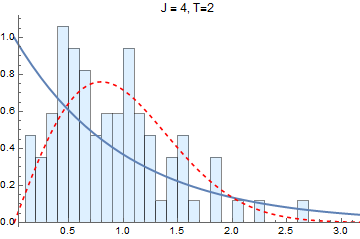}
\includegraphics[scale=0.43]{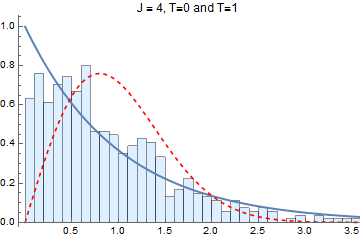}
\includegraphics[scale=0.43]{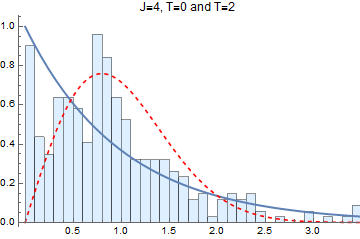}
\includegraphics[scale=0.43]{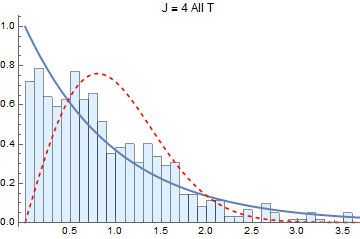}
\end{figure}

\begin{figure}[H]
\caption{NNS histograms for $J=5^+$ with various $T$}
\includegraphics[scale=0.43]{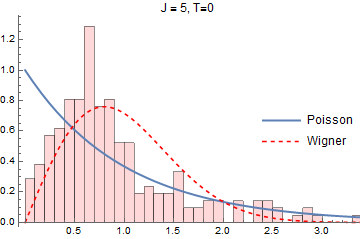}
\includegraphics[scale=0.43]{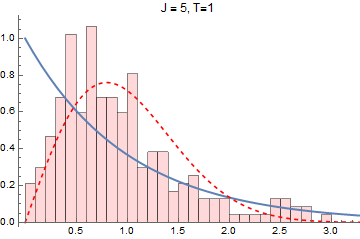}
\includegraphics[scale=0.43]{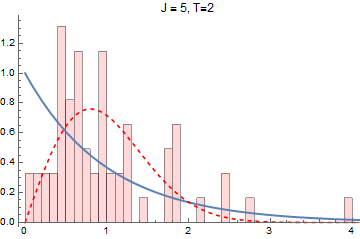}
\includegraphics[scale=0.43]{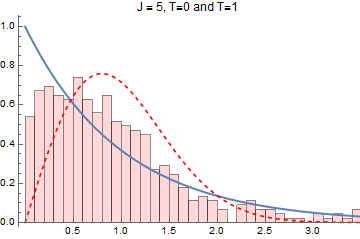}
\includegraphics[scale=0.43]{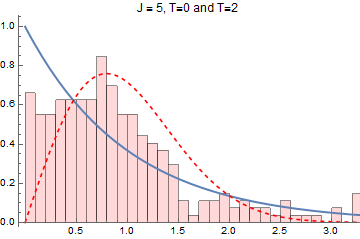}
\includegraphics[scale=0.43]{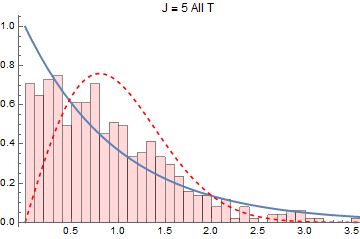}
\end{figure}

\begin{figure}[H]
\centering
\caption{NNS histograms with the Coulomb interaction included}
\includegraphics[scale=0.43]{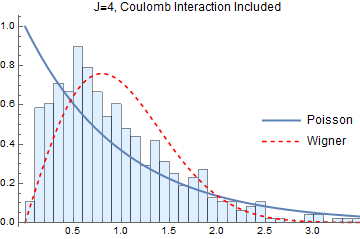}
\includegraphics[scale=0.43]{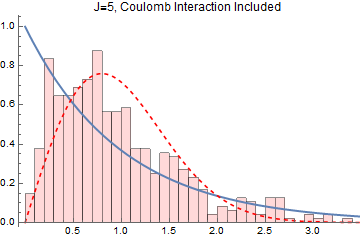}
\end{figure}

\begin{figure}[H]
\centering
\caption{NNS histogram for mixed $J=4^+$ and $J=5^+$ states of any isospin}
\includegraphics[scale=0.43]{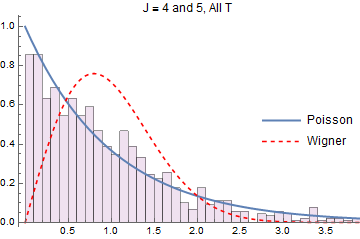}
\end{figure}

Let us first analyze the figures visually. We see that for
$J=4^{+}$ when we include all isospins we get a Poisson-like distribution
with a peak at near-zero spacing. However when we consider the isopins
one by one we get more Wigner looking distributions for all 3 cases--$T=0$
only, $T=1$ only and $T=2$ only. Even though there are much fewer $T=2$
states than $T=0$, when we include both $T=0$ and $T=2$ we again obtain a more Poisson-like distribution. We get very similar results when we
look at $J=5^{+}$ states. When we combine $J=4$ and $J=5$ with all isospins
we get an even more pronounced Poisson behavior. Thus
all of our results are in accord with the concepts of Wigner {[}1{]}.

We next consider the variances of these distributions. The results are listed in Tables I and II. As previously stated, our spacings are expressed in units of the mean spacing for each particular configuration of $J$ and $T$. This normalizes the samples such that their means are 1: Therefore, they can be quantitatively compared to Poisson distributions by comparing their variances to 1, since for Poisson distributions the variance equals the mean.

\begin{table}[H]
\caption{Means and variances for $J=4^+$}
\centering
\begin{tabular}{|c|c|l|l|}

  \hline
  Isospin & Number of Spacings & Mean & Variance (with mean normalized to 1) \\
  \hline \hline 
  0 & 258 & 0.1094089147 & 0.6107983913 \\
  \hline
  1 & 279 & 0.09231182796 & 0.9754159939 \\
  \hline
  2 & 85 & 0.2367364706 & 0.5475221429 \\
  \hline
  0 and 1 & 538 & 0.05151988848 & 1.27191479 \\
  \hline
  0 and 2 & 344 & 0.07839883721 & 0.9897211244 \\
  \hline
  0, 1, and 2 & 624 & 0.04295961538 & 1.187964501 \\
  \hline
  (Coulomb) & 479 & 0.03760041754 & 0.6720058684 \\
  \hline
\end{tabular}
\end{table}

\begin{table}[H]
\caption{Means and variances for $J=5^+$}
\centering
\begin{tabular}{|c|c|l|l|}

  \hline
  Isospin & Number of Spacings & Mean & Variance (with mean normalized to 1) \\
  \hline \hline 
  0 & 210 & 0.1265361905 & 0.8842338025 \\
  \hline
  1 & 235 & 0.1051753191 & 0.8924135702 \\
  \hline
  2 & 61 & 0.2664868852 & 0.5352917635 \\
  \hline
  0 and 1 & 445 & 0.0579132287 & 0.953985486 \\
  \hline
  0 and 2 & 272 & 0.09048088235 & 0.8679745249 \\
  \hline
  0, 1, and 2 & 508 & 0.04859724409 & 1.036806897 \\
  \hline
  (Coulomb) & 479 & 0.04457265136 & 0.5361667558 \\
  \hline
\end{tabular}
\end{table}
 We note for $J=4^+$ and $T=0$ the mean is 0.109, for $J=4^+$ and $T=1$ the mean is 0.092, and for $J=4^+$ with mixed $T=0$ and $T=1$ the mean is 0.0515. The reduction in the mean for the combined $T=0$ and $T=1$ spacings can be understood by the fact that there is no level repulsion  between $T=0$ and $T=1$ states, so they can come close to each other in energy.
When expressed  in terms of the parameter $s$, the ratio of the level spacing to the mean level spacing, the variance for a Poisson distribution  is 1, while for the Wigner distribution it is $4/\pi-1 \approx 0.2732$. This results in table I are consistent with this, in that the smallest variances  correspond to cases where the distributions are closest to the Wigner distributions i.e. cases where all states have the same isospin -- $T=0$ only, $T=1$ only and $T=2$ only. Likewise, the distribution which results from including the Coulomb interaction looks more Wigner-like than the Coulomb-less histogram with all isospins. Indeed, the variance in the Coulomb case is smaller than in the Coulomb-less case.

We note that while removing any states, not just those of a particular isospin or any other property, necessarily replaces spacings adjacent to these levels with larger spacings equal to their sum, the resulting NNS histogram reacts differently to removing states of a certain $T$ that it does to removing random states. We show this in Figure 5, where a histogram is produced when randomly removing a number of states equal to the number of $T=1$ and $T=2$ states in the $J=4^+$ data set.

\begin{figure}[H]
\centering
\caption{Randomly removing states}
\includegraphics[scale=0.43]{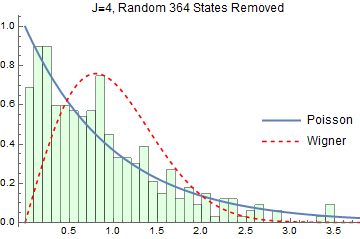}
\end{figure}
 
Note the large portion of spacings in the near-zero region, and thus strong agreement with the Poisson distribution. In comparison, the $J=4^+$ $T=0$ histogram has relatively few spacings in this region, and more closely resembles the Wigner distribution over these small spacing intervals. We therefore conclude that removing states of certain isospin is of physical significance, and that the resulting effect on the probability density can not be likened to removing randomly distributed states. 

\section{Selected Experiments and Closing remarks.}
   The idea that symmetry plays an important  role in energy level spacing distributions  is certainly not new but, as they say, the devil is in the details. Looking at the figures in this work, we can first make a rough analysis. Certainly if we limit ourselves to levels which all have the same isospin we get distributions which look more Wigner-like than Poisson-like, as compared with the case where we include all isospins. If we include states of more than one angular momentum, in this case $J=4^+$ and $J=5^+$, we obtain a distribution which is even closer to a Poisson distribution.
   
  If, however, we take a closer look at the cases where one has only one isospin present, there are some deviations from an ideal Wigner distribution. In each of these single $T$ cases, both for $J=4^+$ and for $J=5^+$, there are still entries near zero level spacing, which indicates some lack of level repulsion. One naturally may question the origin of this - is it due to  some  additional symmetry? We do not have an answer at the present time, but it is a question worth pursuing.
A perhaps more striking feature is that for level spacing beyond the mean, the Poisson curve gives a better fit to the calculated spectra than does Wigner, even in the single $T$ cases. Consistently the Wigner distribution is above the results whereas Poisson gives an excellent fit.

Another point of interest is that when, with all levels included, we include the Coulomb interaction, we get a curve with a deficit of spacings near zero. That is, the curve looks more Wigner-like as compared with the case where the Coulomb interaction  is absent. This can be explained by the fact that isospin is no longer a good quantum number when the Coulomb interaction is included - a given state is a mixture of all 3 isospins. By removing the isospin symmetry, we increase the amount of level repulsion.

We next make  selected some remarks  about  experiments. In the literature one can see evidence of both Poisson and Wigner-like behaviors. Concerning the former, we cite the work of Huizenga and Kasanov [12]. In their analysis they have states of both positive and negative parity i.e. mixed symmetry, so it expected that an exponential distribution would result. In contrast, the work of Garg et al. one has dominantly  $J=\frac{1}{2}^+$  resonant states from the reaction $\mathrm{n}\,+ \, ^{232}$Th up to an energy of 3.9 keV. Perhaps the most important point of this paper is that the shape they obtain is remarkably close to what we get when we consider one isospin. The shape is Wigner-like but the  deviations from Wigner are almost exactly  the same  as what we calculate for a $J=4^+$ $T=0$ i.e. not quite zero at zero spacing and closer to Poisson than Wigner at large spacing.  This suggests a universal  pattern in both experiment and shell model theory. In the selected theoretical single $J$ single $T$ calculations we performed, we get the same pattern no matter what angular momentum or isospin we used. In an experiment which apparently does not involve mixed symmetries, we obtain the same pattern as that which resulted from our calculations. This is perhaps the most important point of this work.  

   But what about isospin in the Garg et al. [13] experiment? In order to obtain a state  with isospin larger than $(N-Z)/2$  in a heavy nucleus with a large neutron excess, one has to excite a proton to a level above the neutron excess. Thus for the most part states of higher isospin lie higher in energy than the states seen in the Garg et al. experiment.

\section{Acknowledgments}
M.Q. was  supported via REU by NSF grant PHY-1560077. A.K. was supported by the Rutgers Aresty summer 2016 internship. We thank Shadow Robinson for his help in installing the Nushellx  program.


\begin{thebibliography}{10}
\bibitem[1]{kat} E. Wigner. On the statistical distribution of the
widths and spacings of nuclear resonance levels. Mathematical Proceedings
of the Cambridge Philisophical Society, pages 790\textendash 798,
1951.

\bibitem{key-1} E. Wigner. Characteristic vectors of bordered matrices
with infinite dimensions. The Annals of Mathematics, 62(3):548\textendash 564,
195

\bibitem{key-3}F.J. Dyson. Statistical theory of the energy levels
of complex systems. i. Journal of Mathematical Physics, 3(1):140\textendash 156,
1962.

\bibitem{key-6}{]} F.J. Dyson. Statistical theory of the energy levels
of complex systems. iii. Journal of Mathematical Physics, 3(1):166\textendash 175,
1962.

\bibitem{key-2}. F. J. Dyson and M. L. Mehta, J. Math. Phys., 4701,
(1963) .

\bibitem{key-7}T. A. Brody, J. Flores, J. P. French, P. A. Mello,
A. Pandey \& S. S. M. Wong, Rev. Mod. Phys. 53, 385,(1981). 

\bibitem{key-5}A. Y. Abul-Magd and H. A. Weidenmüller, Phys. Lett.
162B, 223 ,(1985). {[}5{]}.

\bibitem{key-8}J. F. Shriner, G. E. Mitchell, and T. von Egidy, Z.
Phys. A 338, 309, (1991)

\bibitem{key-9}C. W. Johnson, G. F. Bertsch, D. J. Dean, and I. Talmi
Phys. Rev. C 61, 014311 (1999) 

\bibitem{key-11}K. Van Houcke, S. M. A. Rombouts, K. Heyde, and Y.
Alhassid Phys. Rev. C 79, 024302 (2009) 

\bibitem{key-10}Y. Lu, Y. M. Zhao, and A. Arima Phys. Rev. C 91,
027301 (2015)


\bibitem{key-1}The Shell- Model Code NUSHELLX@MSU , B.A. Brown and
W.D.M. Rae ,  http://www.sciencedirect.com/science/article/pii/S0090375214004748

\bibitem{key-10}
J.R. Huizenga and A.A. Kasanov, Nucl. Phys.A 98,614 (1967).

\bibitem{key-10}
J.B. Garg, J. Rainwater, J.S. Peterson and W.W. Havens Jr., Phys. Rev. 134, B985 (1964)

\end{thebibliography}
\end{document}